%% file: main.tex
\documentclass{article}

\usepackage{arxiv}

\usepackage[utf8]{inputenc} 
\usepackage[T1]{fontenc}    
\usepackage{hyperref}       
\usepackage{url}            
\usepackage{booktabs}       
\usepackage{amsfonts}       
\usepackage{nicefrac}       
\usepackage{microtype}      
\usepackage{lipsum}		
\usepackage{graphicx}
\usepackage{doi}

\usepackage{natbib}
\usepackage{subcaption} 
\usepackage{enumitem}  
\usepackage{float}

\graphicspath{ {./figures/} }

\setlist[itemize]{noitemsep, topsep=0pt, parsep=0pt, partopsep=0pt}

\title{Self-Supervised Pre-Training with Joint-Embedding Predictive Architecture Boosts ECG Classification Performance}

\author{ \href{https://orcid.org/0000-0002-1525-8798}{\includegraphics[scale=0.06]{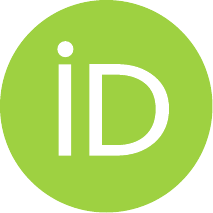}\hspace{1mm}Kuba Weimann} \\
	Zuse Institute Berlin \\
	\texttt{kuba.weimann@zib.de} \\
	\And
	\href{https://orcid.org/0000-0002-5590-5726}{\includegraphics[scale=0.06]{orcid.pdf}\hspace{1mm}Tim O. F. Conrad} \\
	Zuse Institute Berlin \\
	\texttt{conrad@zib.de} \\
}

\date{}

\renewcommand{\shorttitle}{}

\hypersetup{
pdftitle={Self-supervised Pre-training of ECG Classifiers with Joint-Embedding Predictive Architecture},
pdfsubject={cs.LG},
pdfauthor={Kuba Weimann, Tim O. F. Conrad},
pdfkeywords={Self-supervised learning, Joint-embedding predictive architecture, ECG classification},
}

\begin{document}
\maketitle

\begin{abstract}
Accurate diagnosis of heart arrhythmias requires the interpretation of electrocardiograms (ECG), which capture the electrical activity of the heart. Automating this process through machine learning is challenging due to the need for large annotated datasets, which are difficult and costly to collect. To address this issue, transfer learning is often employed, where models are pre-trained on large datasets and fine-tuned for specific ECG classification tasks with limited labeled data. Self-supervised learning has become a widely adopted pre-training method, enabling models to learn meaningful representations from unlabeled datasets. In this work, we explore the joint-embedding predictive architecture (JEPA) for self-supervised learning from ECG data. Unlike invariance-based methods, JEPA does not rely on hand-crafted data augmentations, and unlike generative methods, it predicts latent features rather than reconstructing input data. We create a large unsupervised pre-training dataset by combining ten public ECG databases, amounting to over one million records. We pre-train Vision Transformers using JEPA on this dataset and fine-tune them on various PTB-XL benchmarks. Our results show that JEPA outperforms existing invariance-based and generative approaches, achieving an AUC of 0.945 on the PTB-XL all statements task. JEPA consistently learns the highest quality representations, as demonstrated in linear evaluations, and proves advantageous for pre-training even in the absence of additional data.
\end{abstract}


\section{Introduction}

Self-supervised learning has become a widely adopted method for pre-training deep neural networks on large-scale datasets \citep{kolesnikov2020big, brown2020language, jia2021scaling, radford2023robust, touvron2023llama2, oquab2024dinov}. By deriving the training signal directly from the data, self-supervised learning can leverage massive, unlabeled datasets (e.g., \citep{schuhmann2021laion400, commoncrawl}) for model pre-training. The resulting models tend to capture generalizable features that can be fine-tuned for a diverse range of downstream tasks, facilitating faster convergence and improved performance. This is particularly beneficial in fields where access to labeled data is constrained, either due to high annotation costs or privacy regulations that limit data sharing \citep{rieke2020future}. In the context of electrocardiogram (ECG) classification, self-supervised learning presents a promising alternative to traditional supervised pre-training, improving model generalization to specific medical tasks without the need for explicit annotations.

\subsection{Related Work}

Self-supervised learning typically involves transforming or augmenting data to generate pairs of related samples, which the model is trained to predict or reconstruct. This approach can be categorized into two main types: invariance-based and generative methods.

\paragraph{Invariance-based methods.} Invariance-based methods aim to produce similar embeddings for compatible input pairs and dissimilar embeddings for incompatible pairs. A major challenge with these methods is avoiding representation collapse, where the model produces the same embedding regardless of input variability. Several techniques have been developed to mitigate this issue, such as contrastive methods \citep{oord2019representation, he2020momentum, chen2020simple}, clustering-based approaches \citep{caron2018deep, caron2020unsupervised}, and self-distillation techniques \citep{grill2020bootstrap, chen2021exploring, caron2021emerging}. While these methods, used predominantly in computer vision, can learn highly semantic representations \citep{assran2023the}, they can also introduce biases toward features invariant to specific data augmentations. Such biases may negatively impact downstream tasks, particularly when attempting to generalize augmentations designed for images to other modalities, like ECG data.

\paragraph{Generative methods.} On the other hand, generative methods focus on reconstructing an input $y$ from a transformed version $x$, often using masking \citep{xie2022simmim, he2022masked} to create compatible pairs. Further examples include denoising autoencoders \citep{vincent2008extracting} and variational autoencoders \citep{kingma2013auto}. Unlike invariance-based approaches, generative methods do not require prior knowledge or hand-crafted augmentations of the data. However, they tend to be computationally intensive and can be less efficient at learning representations, as they require modeling relationships within the low-level input space (e.g., pixel-space) \citep{oord2019representation, assran2023self}.

\paragraph{ECG classification.} An electrocardiogram (ECG) records the heart's electrical activity as a time-series waveform (see Figure \ref{figure:ECG}), commonly used to diagnose heart conditions. Training accurate ECG classifiers is particularly challenging due to the high cost of labeling ECG records, which often leads to a scarcity of annotated data. A common strategy to mitigate this challenge is the use of transfer learning, and more specifically, self-supervised learning, where pre-trained models are leveraged to improve performance on specific ECG classification tasks. Traditional pre-training approaches for ECG data have typically relied on supervisory signals from other labeled datasets \citep{kachuee2018ecg, salem2018ecg, strodthoff2021deep}.

\paragraph{Self-supervised learning from ECG data.} Recent studies have increasingly focused on self-supervised learning, utilizing contrastive methods \citep{weimann2021transfer, kiyasseh2021clocs, mehari2022selfsupervised, oh2022leadagnostic, le2023sclst, lai2023practical}, multi-modal data integration \citep{li2024frozen}, knowledge distillation \citep{qin2023mvkt}, and generative methods \citep{sawano2022masked, vaid2023foundational, zhang2023maefe, na2024guiding}. While ongoing research aims to identify effective transformations for ECG data to create compatible pairs \citep{soltanieh2022analysis}, recent findings indicate that invariance-based methods may often underperform in downstream ECG classification tasks \citep{mehari2022selfsupervised, na2024guiding}. Instead, \citet{na2024guiding} advocate for reconstructing spatio-temporally masked segments of the ECG waveform using a masked autoencoder architecture \citep{he2022masked}, while \citet{mehari2022selfsupervised} propose predicting future waveforms in latent space using the Contrastive Predictive Coding (CPC) framework \citep{oord2019representation}. This latter approach currently represents the state-of-the-art for various ECG classification benchmarks.

\subsection{Proposed Approach}

\begin{figure}
\centering
\begin{subfigure}[b]{0.45\columnwidth}
    \centering
    \includegraphics[width=\textwidth]{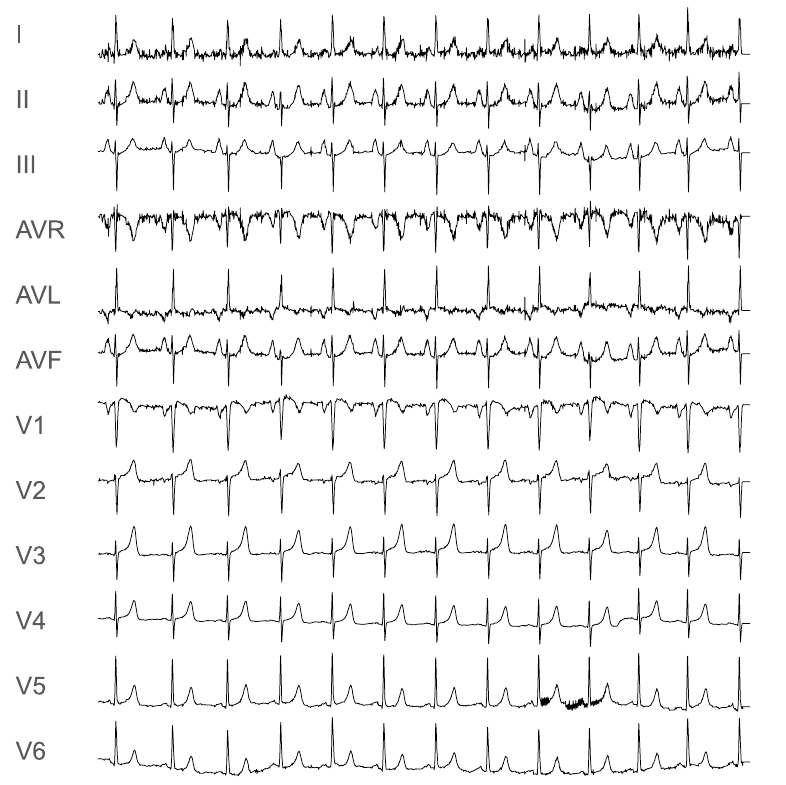}
    \caption{12-lead ECG}
    \label{figure:ECG}
\end{subfigure}
\hspace{0.05\columnwidth}
\begin{subfigure}[b]{0.45\columnwidth}
    \centering
    \includegraphics[width=\textwidth]{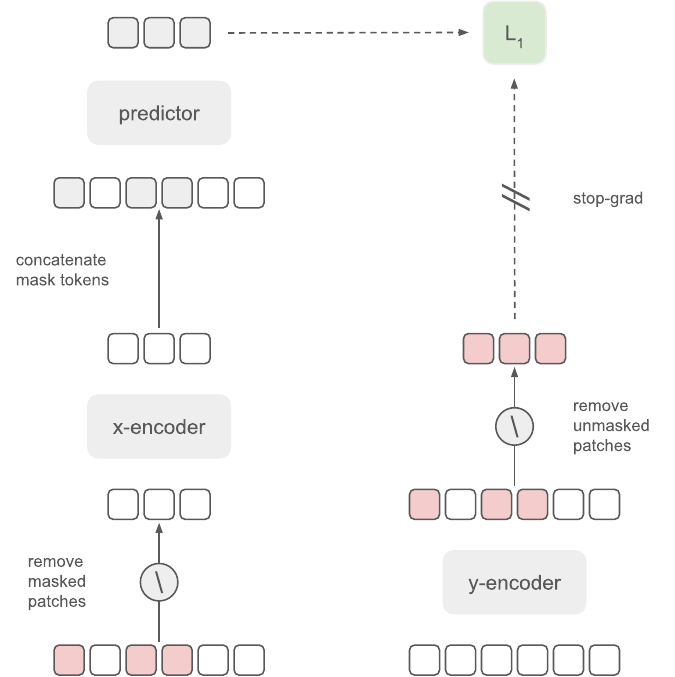}
    \caption{Joint-Embedding Predictive Architecture}
    \label{figure:JEPA-architecture}
\end{subfigure}
\caption{\textbf{Model overview.} A 12-lead ECG $y$ (a) undergoes preprocessing and is divided into patches. Contiguous blocks of patches from $y$ are randomly masked, and the remaining patches are used to form an ECG $x$. Both ECGs are processed by the joint-embedding predictive architecture (JEPA) \citep{assran2023self} for feature prediction in latent space (b). The y-encoder embeds the patches of $y$ to generate target embeddings. Simultaneously, the x-encoder processes the unmasked patches in $x$, which, along with mask tokens indicating the positions of the masked patches, are fed to the predictor. The predictor then outputs patch-level predictions for the masked patches. The model minimizes the $L_1$ reconstruction loss between the target and predicted embeddings. To prevent model collapse, the y-encoder is not directly trained, but rather updated using an exponential moving average of the x-encoder's weights.}
\label{figure:model-overview}
\end{figure}

In this work, we introduce the use of the joint-embedding predictive architecture (JEPA) \citep{assran2023self} for self-supervised learning from ECG data (see Figure \ref{figure:JEPA-architecture}). JEPA addresses key limitations of traditional invariance-based and generative methods. Unlike generative methods, which reconstruct data in the low-level input space, JEPA predicts missing information directly in the latent feature space. This approach allows JEPA to disregard irrelevant low-level details, thereby emphasizing more meaningful semantic features in the learned representations. Additionally, JEPA does not depend on hand-crafted data augmentations to create compatible pairs, making it more adaptable to non-vision modalities, such as time-series data. A detailed explanation of the method is provided in Section \ref{section:method}.

We adopt a standard transfer learning procedure: first, we pre-train Vision Transformers \citep{dosovitskiy2021an} using JEPA, and then fine-tune these models on specific ECG classification tasks with limited labeled data. For unsupervised pre-training, we combine ten publicly available ECG databases \citep{gow2023mimic, ribeiro2021code15, wagner2020ptbxl, reyna2021will}, resulting in a dataset of over one million ECG records (details in Section \ref{section:pretraining-data}). We modify the model architecture for ECG data and implement a simple yet effective masking strategy to create compatible pairs from the ECG waveforms. The code for our implementation is publicly available at \mbox{\url{https://github.com/kweimann/ECG-JEPA}}.

We fine-tune the pre-trained Vision Transformers on multiple ECG classification tasks, using the PTB-XL benchmark \citep{wagner2020ptbxl} to evaluate their performance (see Section \ref{section:results}). Our findings show that JEPA not only excels in linear evaluation, demonstrating the model's ability to capture useful general features, but also achieves state-of-the-art performance when fine-tuned for specific ECG classification tasks. Moreover, our insights from scaling both the dataset and model size during pre-training offer valuable perspectives on how self-supervised learning with JEPA can enhance the analysis of ECG data at scale (see Section \ref{section:conclusion}).

Our main contributions are as follows:

\begin{itemize}
    \item We propose the use of the joint-embedding predictive architecture (JEPA) for self-supervised learning from ECG data. Our findings demonstrate the superior performance of JEPA on several PTB-XL benchmarks. Specifically, on the PTB-XL all statements task, JEPA achieves an AUC of 0.945, surpassing the previously leading \texttt{CPC} model \citep{mehari2022selfsupervised} (0.942 AUC). Additionally, JEPA outperforms \texttt{ST-MEM} \citep{na2024guiding}, with an AUC of 0.935 compared to 0.933 on their respective PTB-XL benchmark.
    
    \item We conduct an extensive evaluation of JEPA's performance across various experimental conditions. JEPA consistently learns the highest quality representations, as demonstrated by linear evaluation on multiple ECG classification tasks. Furthermore, pre-training with JEPA is always advantageous, even when no additional data is available for pre-training. 
    
    \item We create a large-scale dataset for unsupervised pre-training on ECG data. We consolidate ten publicly available ECG datasets, resulting in over a million ECG records from institutions worldwide.
\end{itemize}

\section{Method: Joint-Embedding Predictive Architecture}
\label{section:method}

The joint-embedding predictive architecture (JEPA) \citep{assran2023self, bardes2024revisiting} learns to predict the representation of an ECG $y$ from the representation of a compatible ECG $x$. JEPA incorporates an \textit{encoder} network $E_\theta$, which computes the ECG representations, and a \textit{predictor} network $P_\phi$, which predicts the representation of $y$ from the representation of $x$, conditioned on an additional variable $z$ that describes the transformation between $y$ and $x$ (see Figure \ref{figure:JEPA-architecture}). Our primary objective with JEPA is to learn useful ECG features that can improve the performance of ECG classifiers on downstream tasks, particularly when data availability is limited. In the following, we first describe the training objective, which solves a feature prediction task, and then we describe the key components of the architecture.

\subsection{Training Objective}

The training objective of JEPA is to minimize the reconstruction loss between the actual representation of $y$ and its prediction from $x$, conditioned on the variable $z$:

\begin{equation}
\label{equation:training-objective}
    \mathcal{L}_{\theta, \phi} = \| P_\phi(E_\theta(x), z) - sg(\bar{E}_\theta(y)) \|_1 .
\end{equation}

To avoid a naive solution in which the outputs of the encoder network collapse to a constant representation, a stop-gradient operation, $sg(\cdot)$, is used along with an exponential moving average of the encoder, $\bar{E}_\theta(\cdot)$. For a theoretical explanation of this strategy to prevent model collapse, see \citet{grill2020bootstrap} and \citet{bardes2024revisiting}.

\subsection{Feature Prediction Task}

JEPA learns to predict the representation of ECG $y$ from a compatible ECG $x$. The transformation between $y$ and $x$ involves a masking operation, similar to that used in the Masked Autoencoder (MAE) approach \citep{he2022masked}, where random patches of the input are masked and the missing elements are subsequently reconstructed. However, a key distinction in JEPA is that this reconstruction occurs in the latent feature space rather than directly in the input space. 

To initiate this process, an ECG $y$ is divided into regular, non-overlapping patches. Following \citet{assran2023self}, we sample several contiguous blocks of these ECG patches and then use their complement to form the ECG $x$. The unmasked patches, $x$, are processed by the encoder network $E_\theta$, while the exponential moving average of the encoder, $\bar{E}_\theta$, handles all patches that constitute the full ECG $y$. Subsequently, the patch-level representations of $x$ are fed into the predictor network $P_\phi$ along with a learnable mask token for each masked patch, represented by the variable $z$. The predictor network then generates patch-level predictions for the masked patches, whose actual representations are calculated by $\bar{E}_\theta$. Within this feature prediction task, the training objective, as described in Equation (\ref{equation:training-objective}), is to minimize the $L_1$ loss associated with predicting the patch-level features of the masked patches of an ECG $y$ (see Figure \ref{figure:JEPA-architecture}).

\subsection{Masking}

In our work, the masking transformation that creates compatible pairs operates on the ECG waveform. In contrast to \citet{assran2023self}, who generate 2-dimensional masks for images, or \citet{bardes2024revisiting}, who create 3-dimensional spatio-temporal masks for videos, the 1-dimensional nature of ECG simplifies the process of generating masks. In higher dimensions, masking can be more complicated due to overlaps between the contiguous blocks that are masked. These overlaps must be carefully managed when forming the context $x$ \citep{assran2023self}. Additionally, to ensure training efficiency, a consistent number of patches need to be masked across all samples in a batch to maintain uniform input shapes. 

Our approach addresses the issues of managing overlaps and maintaining consistency in masking by employing a dynamic strategy for handling contiguous blocks of ECG patches. We begin by identifying a list of contiguous blocks of unmasked ECG patches. During masking, we randomly select a block from this list and mask either the entire block or part of it to create a new contiguous masked segment. If the block is only partially masked, the list is updated with the remaining unmasked portions; otherwise, the block is removed. This approach ensures the systematic reduction of unmasked areas while guaranteeing uniform input shapes across the batch without additional heuristics.

Throughout pre-training, we aim to mask between 75\% and 85\% of the patches in any given batch to ensure sufficient difficulty of the prediction task and an adequate context size to solve it. Additionally, we set a minimum block size at 5\% of the patches to prevent overly fragmented data.

\subsection{Model Parameterization}

We parameterize both the encoder and predictor networks in JEPA using Vision Transformers \citep{dosovitskiy2021an}, with the encoder employing a single register \citep{darcet2024vision}. Vision Transformers are a particularly suitable architecture for the feature prediction task because they allow flexible patch-based operations and are highly scalable \citep{he2022masked}. We use three configurations that differ primarily in model dimensionality and depth: \texttt{ViT-B} with 85.2M parameters, \texttt{ViT-S} with 14.3M parameters, and \texttt{ViT-XS} with 3.6M parameters. Detailed descriptions of the architecture and hyperparameters of these models are available in Appendix \ref{appendix:detailed-model-parameterization}.

For tokenizing the ECG waveform, we use a 1-dimensional strided convolution that outputs non-overlapping patches, each of size 25. This sequence of patches (or tokens) is subsequently processed by a stack of transformer blocks. Masking occurs before feeding $x$ into the encoder to establish the context and after feeding $y$ to the exponential moving average of the encoder to generate targets. Drawing inspiration from masked autoencoders \citep{he2022masked}, the predictor concatenates the sequence of $x$-embeddings with a sequence of learnable mask tokens. These tokens incorporate positional encodings \citep{vaswani2017attention} to indicate the temporal positions of the targets.

\begin{table}
\caption{\textbf{ECG databases that constitute the unsupervised pre-training dataset.}}
\label{table:data}
\begin{center}
\begin{scriptsize}
\begin{sc}
\begin{tabular}{lllll}
\toprule
Dataset                     & Records   & ECG Seconds & \vtop{\hbox{\strut Sampling}\hbox{\strut Ratio}} & Source \\
\midrule
MIMIC-IV-ECG                & 800,035   & 8,000,350  & 0.7 & {\normalfont \citet{gow2023mimic}}       \\
CODE-15                     & 128,033   & 1,311,060  & 0.1 & {\normalfont \citet{ribeiro2021code15}}       \\
PTB-XL (training partition) & 17,439    & 174,390    & 0.05 & {\normalfont \citet{wagner2020ptbxl}}       \\
Chapman-Shaoxing            & 10,247    & 102,470    & 0.01875 & {\normalfont \citet{Zheng2020a12lead, reyna2021will}}       \\
CPSC                        & 6,867     & 109,585    & 0.025 & {\normalfont \citet{liu2018open, reyna2021will}}       \\
CPSC-Extra                  & 3,441     & 54,819     & 0.0125 & {\normalfont \citet{liu2018open, reyna2021will}}       \\
Georgia                     & 10,292    & 102,920    & 0.0375 & {\normalfont \citet{reyna2021will}}        \\
Ningbo                      & 34,905    & 349,050    & 0.028125 & {\normalfont \citet{zheng2020optimal, reyna2021will}}       \\
PTB                         & 516       & 57,150     & 0.009375 & {\normalfont \citet{bousseljot1995nutzung, reyna2021will}}       \\
St-Petersburg               & 74        & 133,200   & 0.009375 & {\normalfont \citet{reyna2021will}}       \\
\midrule
Total                       & 1,011,849 & 10,394,994  & 1.0 &       \\
\bottomrule
\end{tabular}
\end{sc}
\end{scriptsize}
\end{center}
\end{table}

\subsection{Pre-training}

For pretraining, we use the AdamW optimizer \citep{loshchilov2018decoupled} over 100,000 steps with a batch size of 2048. We configure the optimizer with $\beta_1=0.9$, $\beta_2=0.99$, $\epsilon=1\mathrm{e}{-6}$, and apply weight decay to linear and convolutional layers that follow a cosine schedule \citep{loshchilov2017sgdr} from 0.01 to 0.1. The learning rate also adopts a cosine schedule, decreasing from $1\mathrm{e}{-3}$ to $1\mathrm{e}{-6}$, complemented by a linear warm-up over the first 10,000 steps. The weights of the exponential moving average of the encoder are updated progressively, with a momentum value increasing linearly from 0.998 to 0.9995. On a single A100-80GB-SXM GPU, pre-training a \texttt{ViT-S} takes about 16 hours.

\subsection{Fine-tuning}

After pre-training with JEPA, we keep the exponential moving average of the encoder while discarding the other components. This encoder is then evaluated on the PTB-XL database \citep{wagner2020ptbxl} across various classification benchmarks. Following \citet{mehari2022selfsupervised}, during training, we randomly crop records to 2.5 seconds and, for evaluation, generate multiple crops per record using a stride of 1.25 seconds. The classification performance is determined by averaging the output probabilities of each crop.

We explore three evaluation procedures: linear evaluation, fine-tuning, and two-stage fine-tuning. The linear evaluation, often referred to as \textit{frozen evaluation}, assesses the quality of learned representations with the model's weights frozen. Here, we employ a linear cross-attention module \citep{vaswani2017attention} that combines the patch representations into a single ECG embedding vector, which is then projected onto class probabilities. For the fine-tuning procedure, we add a single linear classification layer to the register embedding \citep{darcet2024vision} and train the entire model end-to-end. In the two-stage fine-tuning approach, we start with the encoder and cross-attention module trained during linear evaluation, unfreeze all layers, and continue training end-to-end.

The models are trained using the AdamW optimizer \citep{loshchilov2018decoupled} over 5,000 steps with a batch size of 128. Detailed configurations of the hyperparameters for all three evaluation procedures are provided in Appendix \ref{appendix:details-eval-procedures}.

\section{Unsupervised Pre-training Dataset}
\label{section:pretraining-data} 

We have consolidated ten public ECG databases to create an unsupervised pre-training dataset (see Table \ref{table:data}). This combined dataset offers over a million 12-lead ECG records from various institutions worldwide:

\begin{itemize}
    \item \textit{MIMIC-IV-ECG} \citep{gow2023mimic}: Comprises a large collection of patient records from the Beth Israel Deaconess Medical Center in Boston, USA.
    \item \textit{CODE-15} \citep{ribeiro2021code15}: Includes data from various medical centers across Brazil, reflecting a broad patient demographic.
    \item \textit{PTB-XL} \citep{wagner2020ptbxl} and \textit{PTB} \citep{bousseljot1995nutzung, reyna2021will}: These datasets feature ECGs from patients at the Physikalisch-Technische Bundesanstalt (PTB) in Germany, known for capturing a wide range of cardiovascular conditions.
    \item \textit{Chapman-Shaoxing} and \textit{Ningbo} \citep{reyna2021will}: Gathered from patients in Shaoxing \citep{Zheng2020a12lead} and Ningbo \citep{zheng2020optimal} hospitals in China, these datasets provide insights into cardiovascular conditions prevalent in East Asian populations.
    \item \textit{CPSC} and \textit{CPSC-Extra} \citep{reyna2021will}: Originates from the China Physiological Signal Challenge in 2018 \citep{liu2018open}.
    \item \textit{Georgia} \citep{reyna2021will}: Collected from a population in Southeastern United States.
    \item \textit{St-Petersburg} \citep{reyna2021will}: Consists of long Holter ECG records from the St. Petersburg Institute of Cardiological Technics in Russia.
\end{itemize}

To standardize the ECG records from these diverse sources, we applied several preprocessing steps: (1) linear interpolation of NaN values, (2) resampling at a uniform rate of 500Hz, (3) normalization based on the mean and standard deviation computed across each database, (4) clipping values that fall outside a range of five standard deviations, and (5) randomly cropping each record to 10 seconds, excluding any records shorter than this duration. Additionally, we remove baseline wander from the CODE-15 and St-Petersburg records, which we found necessary to ensure stable training.

During pre-training, we form mini-batches by sampling records from the combined dataset using ratios that roughly correlate with the sizes of each database (see Table \ref{table:data}). In our experiments, we pre-train models not only on this combined dataset (denoted as \textsc{\scriptsize all datasets} in the results), but also exclusively on the MIMIC-IV-ECG database (\textsc{\scriptsize only MIMIC-IV-ECG}), which constitutes about 80\% of all records, and exclusively on the training partition of PTB-XL (\textsc{\scriptsize only PTB-XL}).

Since the ECG databases are publicly available, we do not redistribute the dataset directly. However, we provide all necessary source code and detailed instructions for preprocessing, allowing for easy recreation of the dataset.

\section{Results: ECG Classification}
\label{section:results}

We present a performance analysis of Vision Transformers (\texttt{ViT}) \citep{dosovitskiy2021an} pre-trained using the joint-embedding predictive architecture (JEPA) \citep{assran2023self} and fine-tuned across various PTB-XL benchmarks \citep{wagner2020ptbxl}. The PTB-XL benchmarks provide increasing levels of granularity, each corresponding to a different set of annotations. Our findings demonstrate that JEPA consistently surpasses the current state-of-the-art in pre-training methods at different granularity levels. Performance is evaluated using the macro area under the receiver operating curve (macro AUC), the primary metric that is universally reported in related studies. 

We evaluate models through both end-to-end fine-tuning and a linear evaluation protocol, which examines the quality of learned representations via a linear classifier. Each experiment was repeated ten times, using early stopping on the validation set (fold 9). We report the average macro AUC on the test set (fold 10) with the standard deviation in parentheses (0.0xx). Unless otherwise noted, the results reflect the highest performance from either fine-tuning or two-stage fine-tuning.

\begin{table}
\caption{\textbf{Performance (macro AUC) on the PTB-XL all statements.} Reported scores represent the average macro AUC over ten runs with the standard deviation in parentheses (0.0xx). Vision Transformers pre-trained with JEPA outperform all other methods, achieving a 0.945 AUC in end-to-end fine-tuning and a 0.940 AUC in linear evaluation.}
\label{table:results-ptbxl-all}
\begin{center}
\begin{scriptsize}
\begin{sc}
\begin{tabular}{ll|lll|l}
\toprule
Model         & Method                    & \vtop{\hbox{\strut Linear}\hbox{\strut Eval.}}    & Fine-tuned & \vtop{\hbox{\strut Fine-tuned}\hbox{\strut (two-stage)}} & Source \\
\midrule
inception1d   & Random Init               & —         & 0.925(08) & —                 & {\normalfont\citet{strodthoff2021deep}}      \\
xresnet1d50   & Random Init               & 0.721(16) & 0.924(05) & —                 & {\normalfont \citet{mehari2022selfsupervised}}      \\
4FC+2LSTM+2FC & Random Init               & 0.711(07) & 0.932(03) & —                 & {\normalfont \citet{mehari2022selfsupervised}}      \\
ViT-B         & Random Init               & 0.867(05) & 0.837(17) & —                 & {\normalfont This work}      \\
ViT-S         & Random Init               & 0.833(06) & 0.883(04) & —                 & {\normalfont This work}      \\
ViT-XS        & Random Init               & 0.815(10) & 0.911(04) & —                 & {\normalfont This work}      \\
\midrule
xresnet1d50   & SimCLR                    & 0.883(03) & 0.927(03) & —                 & {\normalfont \citet{mehari2022selfsupervised}}      \\
xresnet1d50   & BYOL                      & 0.878(02) & 0.929(02) & —                 & {\normalfont \citet{mehari2022selfsupervised}}      \\
4FC+2LSTM+2FC & CPC                       & 0.927(01) & —         & 0.942(01)     & {\normalfont \citet{mehari2022selfsupervised}}          \\
4FC+2LSTM+2FC & CPC (CinC2020 w/o PTB-XL) & 0.919(01) & —         & 0.940(02)     & {\normalfont \citet{mehari2022selfsupervised}}          \\
\midrule
ViT-B         & JEPA (all datasets)       & 0.935(01)          & 0.936(02)          & 0.940(01)                  & {\normalfont This work}      \\
ViT-S         & JEPA (all datasets)       & 0.938(02)          & 0.937(03)          & \textbf{0.945(01)}                  & {\normalfont This work}      \\
ViT-S         & JEPA (only MIMIC-IV-ECG)  & \textbf{0.940(02)}          & 0.936(03)          & 0.944(01)                  & {\normalfont This work}      \\
ViT-S         & JEPA (only PTB-XL)        & 0.926(02)          & 0.926(02)          & 0.930(01)                  & {\normalfont This work}      \\
ViT-XS        & JEPA (all datasets)       & 0.933(02)          & 0.935(03)          & 0.939(00)                  & {\normalfont This work}      \\
ViT-XS        & JEPA (only MIMIC-IV-ECG)  & 0.933(03)          & 0.934(03)          & 0.943(01)                  & {\normalfont This work}      \\
ViT-XS        & JEPA (only PTB-XL)        & 0.931(02)          & 0.925(03)          & 0.940(01)                  & {\normalfont This work}      \\
\bottomrule
\end{tabular}
\end{sc}
\end{scriptsize}
\end{center}
\end{table}

\subsection{Performance on PTB-XL All Statements}

The first benchmark focuses on the PTB-XL all statements task, defined at the highest level of granularity, where each ECG record is assigned at least one of the 71 labels that cover form, rhythm, and diagnostic statements \citep{wagner2020ptbxl}. We evaluated JEPA against methods reported by \citet{mehari2022selfsupervised}, which include pre-trained residual networks and the \texttt{4FC+2LSTM+2FC} model pre-trained using the Contrastive Predictive Coding (CPC) framework \citep{oord2019representation}, to which we refer as \texttt{CPC}. Their model consists of a mix of fully connected layers and LSTM units \citep{hochreiter1997long}. Notably, the most effective \texttt{CPC} model was pre-trained on the entire PTB-XL dataset, while we restricted our pre-training to only the training partition of PTB-XL.

\paragraph{Randomly initialized Vision Transformers underperform compared to residual networks.} Among the randomly initialized Vision Transformers, the smallest model, \texttt{ViT-XS}, achieves the highest AUC of 0.911, yet it still underperforms compared to the \texttt{inception1d} model \citep{ismailfawaz2020inceptiontime}, which scores 0.925 AUC (see Table \ref{table:results-ptbxl-all}). Generally, as the size of a randomly initialized \texttt{ViT} increases, its convergence slows and its performance on the test set deteriorates. Notably, \texttt{ViT-B} exhibits a higher AUC of 0.867 in linear evaluation compared to 0.837 AUC after end-to-end training, demonstrating its high capacity as a feature extractor, and at the same time exhibiting convergence issues on smaller datasets.

\paragraph{JEPA pre-training on PTB-XL without additional data leads to large improvements.} The \texttt{ViT-XS} model, pre-trained with JEPA solely on the training partition of the PTB-XL database (i.e., no additional data), outperforms both randomly initialized (0.925 AUC) and pre-trained residual networks (0.929 AUC), achieving an AUC of 0.940 (see Table \ref{table:results-ptbxl-all}). The larger \texttt{ViT-S} model performs worse in end-to-end fine-tuning, with an AUC of 0.930. Moreover, \texttt{ViT-XS} nearly matches the performance of the \texttt{CPC} model (0.942 AUC) and exceeds it during linear evaluation (0.931 AUC vs. 0.927 AUC), demonstrating that it is always beneficial to pre-train a model, even if no additional data is available.

\paragraph{JEPA outperforms other methods when using all datasets for pre-training.} Scaling the amount of data used for pre-training improves the performance of all Vision Transformers (see Table \ref{table:results-ptbxl-all}). \texttt{ViT-S}, trained on all available ECG datasets, achieves an AUC of 0.945, surpassing the top-performing baseline model, \texttt{CPC}, which has an AUC of 0.942. We observe large improvements with an AUC of 0.944 from pre-training solely on the MIMIC-IV-ECG dataset, which comprises about 80\% of the total data. The inclusion of the remaining nine ECG databases in the pre-training appears to have a negligible impact on performance. Moreover, increasing the size of the model beyond a certain capacity does not yield benefits; in fact, \texttt{ViT-B}, the largest model, records a lower AUC of 0.940 compared to \texttt{ViT-S} (0.945 AUC).

\paragraph{JEPA learns the best representations for the downstream task.} The highest quality of learned representations is evident in Vision Transformers pre-trained with JEPA, as confirmed by the linear evaluation results (see Table \ref{table:results-ptbxl-all}). \texttt{ViT-S} registers an AUC of 0.940, outperforming both the residual networks (0.883 AUC) and \texttt{CPC} (0.927 AUC). These findings support the claim that JEPA effectively develops robust off-the-shelf representations \citep{assran2023self}. Interestingly, fine-tuning the pre-trained \texttt{ViT-S} yields slightly inferior performance (0.936 AUC) compared to using a linear head atop frozen representations (0.940 AUC). Superior performance (0.944 AUC) is only achieved using a two-stage fine-tuning approach, suggesting potential overfitting or suboptimal fine-tuning procedures, including the chosen hyperparameters.

\begin{figure}
\centering
\begin{subfigure}[b]{0.45\columnwidth}
    \centering
    \includegraphics[width=\textwidth]{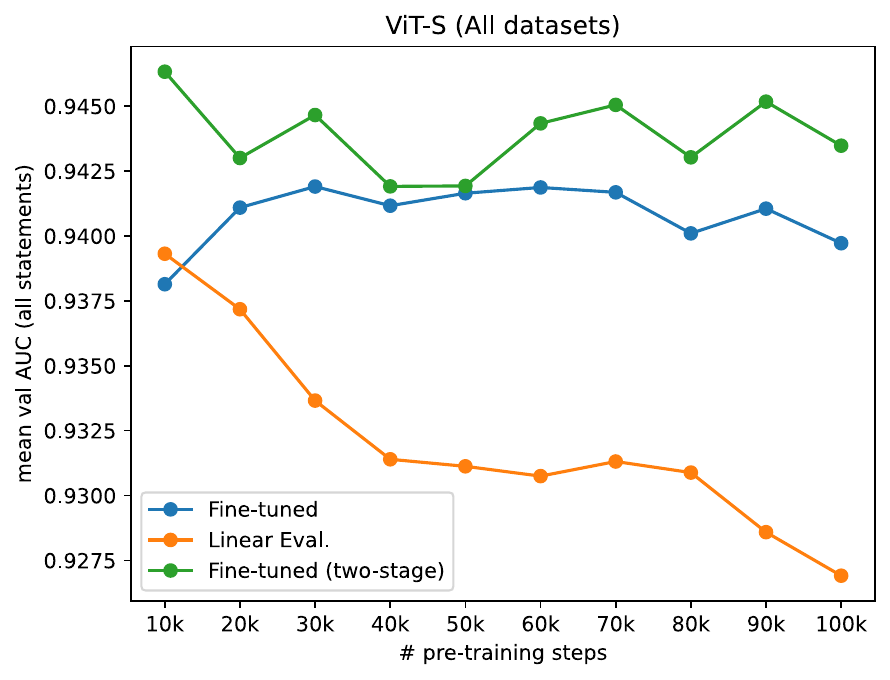}
    \caption{All Statements}
    \label{figure:vits-all-performance}
\end{subfigure}
\hspace{0.05\columnwidth}
\begin{subfigure}[b]{0.45\columnwidth}
    \centering
    \includegraphics[width=\textwidth]{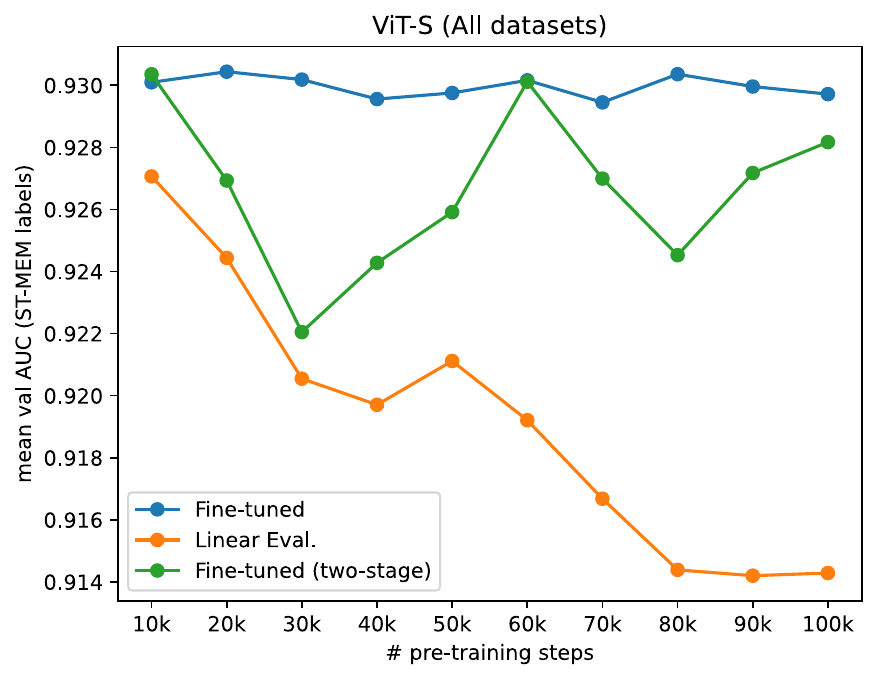}
    \caption{ST-MEM Labels}
    \label{figure:vits-stmem-performance}
\end{subfigure}
\caption{\textbf{Validation performance of \texttt{ViT-S} on PTB-XL downstream tasks.} Higher proficiency at the pre-training objective does not necessarily translate to improved validation performance on downstream tasks.}
\label{figure:vits-performance}
\end{figure}

\paragraph{Proficiency at the pre-training objective does not guarantee improved downstream performance.} Vision Transformers pre-trained with JEPA often achieve their best validation scores before the completion of pre-training (see Figure \ref{figure:vits-performance}). Although the loss gradually decreases throughout the training schedule (not shown here), the \texttt{ViT-S} checkpoint with the highest validation scores is collected after 10,000 steps for linear evaluation, and after 30,000 steps for end-to-end fine-tuning (see Figure \ref{figure:vits-all-performance}). Despite this, the performance on the downstream task during fine-tuning remains stable across the pre-training schedule, while scores from linear evaluation tend to decline over time.

\begin{table}
\caption{\textbf{Performance (macro AUC) on the PTB-XL ST-MEM labels.} Reported scores are the average macro AUC over ten runs with the standard deviation in parentheses (0.0xx). Vision Transformers pre-trained with JEPA outperform all models with a 0.935 AUC in end-to-end fine-tuning and a 0.928 AUC in linear evaluation.}
\label{table:results-ptbxl-stmem}
\begin{center}
\begin{scriptsize}
\begin{sc}
\begin{tabular}{ll|lll|l}
\toprule
Model         & Method                    & \vtop{\hbox{\strut Linear}\hbox{\strut Eval.}}    & Fine-tuned & \vtop{\hbox{\strut Fine-tuned}\hbox{\strut (two-stage)}} & Source \\
\midrule
ViT-B         & MoCo v3                  & 0.739(06) & 0.913(02) & —                        & {\normalfont \citet{na2024guiding}}       \\
ViT-B         & CMSC                     & 0.797(38) & 0.877(03) & —                        & {\normalfont \citet{na2024guiding}}       \\
ViT-B         & MTAE                     & 0.807(06) & 0.910(01) & —                        & {\normalfont \citet{na2024guiding}}       \\
ViT-B         & MTAE+RLM                 & 0.806(05) & 0.911(04) & —                        & {\normalfont \citet{na2024guiding}}        \\
ViT-B         & MLAE                     & 0.779(08) & 0.915(01) & —                        & {\normalfont \citet{na2024guiding}}       \\
ViT-B         & ST-MEM                   & 0.838(11) & 0.933(03) & —                        & {\normalfont \citet{na2024guiding}}       \\
4FC+2LSTM+2FC & CPC                      & —         & 0.934(02) & —                        & {\normalfont \citet{na2024guiding}}        \\
\midrule
ViT-B         & JEPA (all datasets)      & 0.920(02)          & 0.928(03)          & 0.920(02)                         & {\normalfont This work}       \\
ViT-S         & JEPA (all datasets)      & \textbf{0.928(03)}          & 0.933(03)          & \textbf{0.935(02)}                         & {\normalfont This work}       \\
ViT-S         & JEPA (only MIMIC-IV-ECG) & 0.921(03)          & 0.931(03)          & 0.932(02)                         & {\normalfont This work}       \\
ViT-S         & JEPA (only PTB-XL)       & 0.917(01)          & 0.929(02)           & 0.919(01)                         & {\normalfont This work}       \\
ViT-XS        & JEPA (all datasets)      & 0.924(02)          & 0.929(03)          & 0.930(01)                         & {\normalfont This work}       \\
ViT-XS        & JEPA (only MIMIC-IV-ECG) & 0.920(02)          & 0.928(02)          & 0.927(01)                         & {\normalfont This work}       \\
ViT-XS        & JEPA (only PTB-XL)       & 0.919(02)          & 0.928(02)          & 0.925(02)                         & {\normalfont This work}      \\
\bottomrule
\end{tabular}
\end{sc}
\end{scriptsize}
\end{center}
\end{table}

\subsection{Performance on PTB-XL ST-MEM Labels}

We compare JEPA to \texttt{ST-MEM} \citep{na2024guiding}, a Vision Transformer that employs Masked Autoencoder (MAE) \citep{he2022masked} for reconstructing spatio-temporally masked portions of the ECG waveform. Our evaluation follows the labeling scheme used by \citet{na2024guiding}, which uses the PTB-XL diagnostic superclasses \citep{wagner2020ptbxl}---the lowest level of granularity---and excludes all ECG records containing multiple labels. The results presented here should be considered an approximate comparison of the methods, as \texttt{ST-MEM} was trained and evaluated using a custom data split.

\paragraph{JEPA outperforms other methods on the PTB-XL ST-MEM task.} \texttt{ViT-S} pre-trained with JEPA achieves an AUC of 0.935 (see Table \ref{table:results-ptbxl-stmem}), surpassing both \texttt{ST-MEM} with 0.933 AUC and \texttt{CPC} at 0.934 AUC. The validation behavior of Vision Transformers pre-trained with JEPA mirrors that observed in the PTB-XL all statements task (see Figure \ref{figure:vits-performance}): the best validation checkpoint is achieved after 20,000 steps for end-to-end fine-tuning and after 10,000 steps for linear evaluation.

\paragraph{JEPA demonstrates superior performance in linear evaluation.} Using the linear evaluation protocol to assess the quality of learned representations, JEPA pre-training clearly outperforms other methods (see Table \ref{table:results-ptbxl-stmem}). \texttt{ViT-S} records an AUC of 0.928, surpassing \texttt{ST-MEM} at 0.838 AUC and outperforming all other self-supervised learning methods reported by \citet{na2024guiding}.

\subsection{Comprehensive Performance Across PTB-XL Label Hierarchies}

We further assess the performance of Vision Transformers pre-trained with JEPA after fine-tuning on the remaining PTB-XL tasks, which cover a broader range of hierarchically structured ECG labels \citep{wagner2020ptbxl} (the detailed results can be found in Appendix \ref{appendix:additional-results}). For comparison, we use the best performing randomly initialized residual network models reported by \citet{strodthoff2021deep}, omitting scores for randomly initialized Vision Transformers, which consistently underperform relative to residual networks.

\paragraph{JEPA outperforms randomly initialized residual networks in most PTB-XL tasks.} In the diagnostic statements category (see Table \ref{table:results-ptbxl-diagnostic}), \texttt{ViT-S} achieves an AUC of 0.951 after fine-tuning, and \texttt{ViT-B} scores 0.946 AUC in linear evaluation, both surpassing the end-to-end trained \texttt{xresnet1d101} \citep{he2016deep, he2019bag} (0.937 AUC). In diagnostic subclasses (see Table \ref{table:results-ptbxl-subdiagnostic}), \texttt{ViT-S} scores 0.941 AUC after fine-tuning and 0.936 AUC in linear evaluation, outperforming the end-to-end trained \texttt{inception1d} \citep{ismailfawaz2020inceptiontime} (0.930 AUC). In diagnostic superclasses (see Table \ref{table:results-ptbxl-superdiagnostic}), \texttt{resnet1d\_wang} \citep{wang2017time} scores higher (0.930 AUC) than \texttt{ViT-S} (0.928 AUC). On form statements (see Table \ref{table:results-ptbxl-form}), \texttt{ViT-S} records an AUC of 0.911 after fine-tuning and 0.899 AUC in linear evaluation, surpassing the end-to-end trained \texttt{inception1d} (0.899 AUC). Finally, in the rhythm statements category (see Table \ref{table:results-ptbxl-rhythm}), \texttt{ViT-S} achieves an AUC of 0.968 after fine-tuning, matching \texttt{ViT-XS}'s performance in linear evaluation (0.968 AUC), and outperforming the end-to-end trained \texttt{xresnet1d101} (0.957 AUC). Notably, in several instances, the representations learned with JEPA outperform those learned by residual networks in end-to-end training.

\section{Conclusion}
\label{section:conclusion}

In this work, we explored the joint-embedding predictive architecture (JEPA) for self-supervised learning from ECG data. We created an unsupervised pre-training dataset from ten public ECG databases, significantly scaling the data volume used for pre-training compared to many related efforts. We pre-trained several Vision Transformers, varying in embedding sizes and the number of transformer blocks. Subsequently, these models were fine-tuned for ECG classification on various PTB-XL benchmarks, which are hierarchically organized to support different levels of ECG interpretation. Within this diverse benchmarking environment, we evaluated JEPA against several pre-training methods from related studies that employ both invariance-based and generative approaches.

Our results demonstrated that it is always beneficial to pre-train a model, even when additional data is not available. Vision Transformers pre-trained with JEPA on just the training partition of PTB-XL occasionally surpassed methods utilizing much larger volumes of pre-training data. Extending the pre-training data to a larger, related database (MIMIC-IV-ECG) yielded models that outperformed comparable methods on the PTB-XL all statements task. This demonstrates the potential for institutions with access to large, unlabeled ECG databases to significantly enhance downstream performance through pre-training, even if the data originates from a single source. Interestingly, expanding the diversity of our unsupervised pre-training dataset by incorporating more ECG databases had only a marginal impact on performance. Furthermore, our experiments indicated that smaller models, such as \texttt{ViT-XS} and \texttt{ViT-S}, are adequate to achieve the top-tier classification performance, whereas a larger model, \texttt{ViT-B}, did not perform as well, suggesting that even more data could be required to fully utilize the capabilities of larger models.

Throughout our experiments, Vision Transformers pre-trained with JEPA consistently outperformed related methods. On the PTB-XL all statements task, JEPA achieved an AUC of 0.945, surpassing the previous best model, \texttt{CPC}, which scored an AUC of 0.942. On the PTB-XL diagnostic superclasses, with multi-label records excluded, JEPA outperformed both \texttt{CPC} (0.934 AUC) and \texttt{ST-MEM} (0.933 AUC). Furthermore, JEPA demonstrated superior ability in learning high-quality representations, significantly outperforming other approaches in linear evaluations. These findings highlight JEPA's versatility and efficacy for self-supervised learning, extending from images and video to ECG data.

\bibliography{references}


\include{appendix.tex}

\end{document}

%% file: appendix.tex
\appendix

\renewcommand{\thesubsection}{\Alph{subsection}}

\subsection{Detailed Model Parameterization}
\label{appendix:detailed-model-parameterization}

The encoder and predictor are parameterized using Vision Transformers (\texttt{ViT}) \citep{dosovitskiy2021an}, with the encoder employing a single register \citep{darcet2024vision}. The encoder follows the standard \texttt{ViT} architecture, the predictor is a narrow transformer using half the embedding size of the encoder. The detailed hyperparameter values are shown in Table \ref{table:model-parameterization}.

\begin{table}[H]
\caption{\textbf{Model parameterization.}}
\label{table:model-parameterization}
\begin{center}
\begin{scriptsize}
\begin{tabular}{l|lll}
\toprule
Hyperparameter            & \texttt{ViT-XS} & \texttt{ViT-S} & \texttt{ViT-B} \\
\midrule
encoder embedding size    & 192    & 384   & 768   \\
encoder blocks         & 8      & 8     & 12    \\
encoder attention heads   & 4      & 6     & 12    \\
predictor embedding size  & 96     & 192   & 384   \\
predictor blocks       & 8      & 8     & 12    \\
predictor attention heads & 4      & 6     & 12    \\
MLP expansion ratio       & 4.     & 4.    & 4.    \\
dropout                   & 0.     & 0.    & 0.    \\
bias                      & False  & False & False \\
\midrule
{\sc total parameters}          & 3.6M   & 14.3M & 85.2M \\
\bottomrule
\end{tabular}
\end{scriptsize}
\end{center}
\end{table}

\subsection{Details of Evaluation Procedures}
\label{appendix:details-eval-procedures}

All three evaluation procedures use the AdamW optimizer \citep{loshchilov2018decoupled} over 5,000 steps with a learning rate that follows a cosine schedule \citep{loshchilov2017sgdr} with a linear warmup. We use an uniform batch size of 128. Additionally, we use different dropout probabilities depending on the evaluation procedure. The detailed hyperparameter values are shown in Table \ref{table:evaluation-hyperparameters}.

\begin{table}[H]
\caption{\textbf{Evaluation hyperparameters.}}
\label{table:evaluation-hyperparameters}
\begin{center}
\begin{scriptsize}
\begin{tabular}{l|lll}
\toprule
Hyperparameter         & \vtop{\hbox{\strut Linear}\hbox{\strut Eval.}}    & Fine-tuned & \vtop{\hbox{\strut Fine-tuned}\hbox{\strut (two-stage)}} \\
\midrule
learning rate       & 1.0e-3       & 1.0e-3      & 1.0e-4                 \\
final learning rate & 1.0e-3       & 1.0e-5      & 1.0e-6                 \\
warmup steps        & 0            & 200         & 200                    \\
weight decay        & 0.1          & 0.01        & 0.                     \\
AdamW betas         & (0.9, 0.999) & (0.9, 0.99) & (0.9, 0.999)           \\
attention pooling   & True         & False       & True                   \\
dropout             & 0.           & 0.05        & 0.1                    \\
bias                & False        & False       & False                
  \\
\bottomrule
\end{tabular}
\end{scriptsize}
\end{center}
\end{table}

\subsection{Additional Results}
\label{appendix:additional-results}

We report the performance of Vision Transformers pre-trained with JEPA after fine-tuning on various PTB-XL tasks: diagnostic statements (Table \ref{table:results-ptbxl-diagnostic}), diagnostic sublcasses (Table \ref{table:results-ptbxl-subdiagnostic}), diagnostic superclasses (Table \ref{table:results-ptbxl-superdiagnostic}), form statements (Table \ref{table:results-ptbxl-form}), and rhythm statements (Table \ref{table:results-ptbxl-rhythm}). The labels for these tasks are hierarchically organized to facilitate various levels of ECG interpretation \citep{wagner2020ptbxl}. The performance scores are compared to the best randomly initialized residual networks reported by \citet{strodthoff2021deep}. JEPA outperforms the baseline models on every task except on diagnostic superclasses. In several cases, the representations learned with JEPA (based on the linear evaluation) outperform those learned by residual networks in end-to-end training.

\clearpage

\begin{table}[H]
\caption{\textbf{Performance (macro AUC) on the PTB-XL diagnostic statements.}}
\label{table:results-ptbxl-diagnostic}
\begin{center}
\begin{scriptsize}
\begin{sc}
\begin{tabular}{ll|lll}
\toprule
Model         & Method                    & \vtop{\hbox{\strut Linear}\hbox{\strut Eval.}}    & Fine-tuned & \vtop{\hbox{\strut Fine-tuned}\hbox{\strut (two-stage)}} \\
\midrule
xresnet1d101	 & Random Init {\normalfont \citep{strodthoff2021deep}}             & —      & 0.937(08)          & —                        \\
ViT-B           & Random Init              & —       & 0.858(07)          & —                         \\
ViT-S           & Random Init              & —       & 0.903(05)          & —                         \\
ViT-XS          & Random Init              & —       & 0.916(05)          & —                         \\
\midrule
ViT-B           & JEPA (all datasets)      & \textbf{0.946(01)}       & 0.941(04)          & 0.949(01)                         \\
ViT-S           & JEPA (all datasets)      & 0.944(02)       & 0.941(03)          & \textbf{0.951(01)}                         \\
ViT-S           & JEPA (only MIMIC-IV-ECG) & 0.943(03)       & 0.941(03)          & 0.948(01)                         \\
ViT-S           & JEPA (only PTB-XL)       & 0.939(03)       & 0.932(03)          & 0.943(01)                         \\
ViT-XS          & JEPA (all datasets)      & 0.940(03)       & 0.938(04)          & 0.949(01)                         \\
ViT-XS          & JEPA (only MIMIC-IV-ECG) & 0.941(02)       & 0.938(04)          & 0.950(01)                         \\
ViT-XS          & JEPA (only PTB-XL)       & 0.934(03)       & 0.933(02)          & 0.943(02)                        \\
\bottomrule
\end{tabular}
\end{sc}
\end{scriptsize}
\end{center}
\end{table}

\begin{table}[H]
\caption{\textbf{Performance (macro AUC) on the PTB-XL diagnostic subclasses.}}
\label{table:results-ptbxl-subdiagnostic}
\begin{center}
\begin{scriptsize}
\begin{sc}
\begin{tabular}{ll|lll}
\toprule
Model         & Method                    & \vtop{\hbox{\strut Linear}\hbox{\strut Eval.}}    & Fine-tuned & \vtop{\hbox{\strut Fine-tuned}\hbox{\strut (two-stage)}} \\
\midrule
inception1d	 & Random Init {\normalfont \citep{strodthoff2021deep}}             & —      & 0.930(10)          & —                        \\
ViT-B           & Random Init              & —       & 0.851(08)          & —                         \\
ViT-S           & Random Init              & —       & 0.910(06)          & —                         \\
ViT-XS          & Random Init              & —       & 0.924(05)          & —                         \\
\midrule
ViT-B           & JEPA (all datasets)      & 0.934(02)       & 0.938(03)          & 0.939(02)                         \\
ViT-S           & JEPA (all datasets)      & \textbf{0.936(04)}       & 0.939(02)          & 0.935(03)                         \\
ViT-S           & JEPA (only MIMIC-IV-ECG) & 0.935(02)       & \textbf{0.941(03)}          & 0.936(02)                         \\
ViT-S           & JEPA (only PTB-XL)       & 0.930(02)       & 0.926(04)          & 0.937(00)                         \\
ViT-XS          & JEPA (all datasets)      & 0.930(04)       & 0.936(03)          & 0.939(01)                         \\
ViT-XS          & JEPA (only MIMIC-IV-ECG) & 0.929(03)       & 0.937(03)          & \textbf{0.941(01)}                         \\
ViT-XS          & JEPA (only PTB-XL)       & 0.926(02)       & 0.931(03)          & 0.935(02)                        \\
\bottomrule
\end{tabular}
\end{sc}
\end{scriptsize}
\end{center}
\end{table}

\begin{table}[H]
\caption{\textbf{Performance (macro AUC) on the PTB-XL diagnostic superclasses.}}
\label{table:results-ptbxl-superdiagnostic}
\begin{center}
\begin{scriptsize}
\begin{sc}
\begin{tabular}{ll|lll}
\toprule
Model         & Method                    & \vtop{\hbox{\strut Linear}\hbox{\strut Eval.}}    & Fine-tuned & \vtop{\hbox{\strut Fine-tuned}\hbox{\strut (two-stage)}} \\
\midrule
resnet1d\_wang	 & Random Init {\normalfont \citep{strodthoff2021deep}}             & —      & \textbf{0.930(05)}          & —                        \\
ViT-B           & Random Init              & —       & 0.862(06)          & —                         \\
ViT-S           & Random Init              & —       & 0.916(02)          & —                         \\
ViT-XS          & Random Init              & —       & 0.920(01)          & —                         \\
\midrule
ViT-B           & JEPA (all datasets)      & 0.919(01)       & 0.927(02)          & 0.925(01)                         \\
ViT-S           & JEPA (all datasets)      & \textbf{0.920(01)}       & 0.928(01)          & 0.927(01)                         \\
ViT-S           & JEPA (only MIMIC-IV-ECG) & \textbf{0.920(01)}       & 0.928(01)          & 0.928(01)                         \\
ViT-S           & JEPA (only PTB-XL)       & 0.915(01)       & 0.923(01)          & 0.919(01)                         \\
ViT-XS          & JEPA (all datasets)      & 0.915(01)       & 0.926(02)          & 0.926(01)                         \\
ViT-XS          & JEPA (only MIMIC-IV-ECG) & 0.915(01)       & 0.927(01)          & 0.927(01)                         \\
ViT-XS          & JEPA (only PTB-XL)       & 0.912(01)       & 0.924(01)          & 0.921(01)                        \\
\bottomrule
\end{tabular}
\end{sc}
\end{scriptsize}
\end{center}
\end{table}

\clearpage

\begin{table}[H]
\caption{\textbf{Performance (macro AUC) on the PTB-XL form statements.}}
\label{table:results-ptbxl-form}
\begin{center}
\begin{scriptsize}
\begin{sc}
\begin{tabular}{ll|lll}
\toprule
Model         & Method                    & \vtop{\hbox{\strut Linear}\hbox{\strut Eval.}}    & Fine-tuned & \vtop{\hbox{\strut Fine-tuned}\hbox{\strut (two-stage)}} \\
\midrule
inception1d	 & Random Init {\normalfont \citep{strodthoff2021deep}}             & —      & 0.899(22)          & —                        \\
ViT-B           & Random Init              & —       & 0.773(12)          & —                         \\
ViT-S           & Random Init              & —       & 0.822(08)         & —                         \\
ViT-XS          & Random Init              & —       & 0.849(07)          & —                         \\
\midrule
ViT-B           & JEPA (all datasets)      & 0.888(04)       & 0.893(07)          & 0.898(03)                         \\
ViT-S           & JEPA (all datasets)      & 0.896(07)       & 0.894(06)           & \textbf{0.911(03)}                         \\
ViT-S           & JEPA (only MIMIC-IV-ECG) & \textbf{0.899(03)}       & 0.885(06)          & 0.904(02)                         \\
ViT-S           & JEPA (only PTB-XL)       & 0.873(08)       & 0.864(05)          & 0.869(05)                         \\
ViT-XS          & JEPA (all datasets)      & 0.893(04)       & 0.883(10)          & 0.903(02)                         \\
ViT-XS          & JEPA (only MIMIC-IV-ECG) & 0.894(03)       & 0.889(09)          & 0.904(03)                         \\
ViT-XS          & JEPA (only PTB-XL)       & 0.888(04)       & 0.865(09)          & 0.896(03)                        \\
\bottomrule
\end{tabular}
\end{sc}
\end{scriptsize}
\end{center}
\end{table}

\begin{table}[H]
\caption{\textbf{Performance (macro AUC) on the PTB-XL rhythm statements.}}
\label{table:results-ptbxl-rhythm}
\begin{center}
\begin{scriptsize}
\begin{sc}
\begin{tabular}{ll|lll}
\toprule
Model         & Method                    & \vtop{\hbox{\strut Linear}\hbox{\strut Eval.}}    & Fine-tuned & \vtop{\hbox{\strut Fine-tuned}\hbox{\strut (two-stage)}} \\
\midrule
xresnet1d101  & Random Init {\normalfont \citep{strodthoff2021deep}}             & —      & 0.957(19)          & —                        \\
ViT-B           & Random Init              & —       & 0.804(36)          & —                         \\
ViT-S           & Random Init              & —       & 0.836(17)         & —                         \\
ViT-XS          & Random Init              & —       & 0.897(12)          & —                         \\
\midrule
ViT-B           & JEPA (all datasets)      & 0.961(02)       & 0.950(03)          & 0.962(01)                         \\
ViT-S           & JEPA (all datasets)      & 0.967(01)       & 0.956(03)          & 0.966(03)                         \\
ViT-S           & JEPA (only MIMIC-IV-ECG) & 0.966(02)       & 0.952(03)          & \textbf{0.968(01)}                         \\
ViT-S           & JEPA (only PTB-XL)       & 0.931(04)       & 0.949(03)          & 0.934(03)                         \\
ViT-XS          & JEPA (all datasets)      & 0.964(01)       & 0.958(03)          & 0.965(01)                         \\
ViT-XS          & JEPA (only MIMIC-IV-ECG) & \textbf{0.968(02)}       & 0.956(03)          & 0.966(01)                         \\
ViT-XS          & JEPA (only PTB-XL)       & 0.946(03)       & 0.949(06)          & 0.939(04)                        \\
\bottomrule
\end{tabular}
\end{sc}
\end{scriptsize}
\end{center}
\end{table}

%% file: main.bbl
\begin{thebibliography}{58}
\providecommand{\natexlab}[1]{#1}
\providecommand{\url}[1]{\texttt{#1}}
\expandafter\ifx\csname urlstyle\endcsname\relax
  \providecommand{\doi}[1]{doi: #1}\else
  \providecommand{\doi}{doi: \begingroup \urlstyle{rm}\Url}\fi

\bibitem[Assran et~al.(2023{\natexlab{a}})Assran, Duval, Misra, Bojanowski, Vincent, Rabbat, LeCun, and Ballas]{assran2023self}
Mahmoud Assran, Quentin Duval, Ishan Misra, Piotr Bojanowski, Pascal Vincent, Michael Rabbat, Yann LeCun, and Nicolas Ballas.
\newblock Self-Supervised Learning From Images With a Joint-Embedding Predictive Architecture.
\newblock In \emph{Proceedings of the IEEE/CVF Conference on Computer Vision and Pattern Recognition (CVPR)}, pages 15619--15629, 2023{\natexlab{a}}.

\bibitem[Assran et~al.(2023{\natexlab{b}})Assran, Balestriero, Duval, Bordes, Misra, Bojanowski, Vincent, Rabbat, and Ballas]{assran2023the}
Mido Assran, Randall Balestriero, Quentin Duval, Florian Bordes, Ishan Misra, Piotr Bojanowski, Pascal Vincent, Michael Rabbat, and Nicolas Ballas.
\newblock The hidden uniform cluster prior in self-supervised learning.
\newblock In \emph{The Eleventh International Conference on Learning Representations}, 2023{\natexlab{b}}.

\bibitem[Bardes et~al.(2024)Bardes, Garrido, Ponce, Chen, Rabbat, LeCun, Assran, and Ballas]{bardes2024revisiting}
Adrien Bardes, Quentin Garrido, Jean Ponce, Xinlei Chen, Michael Rabbat, Yann LeCun, Mahmoud Assran, and Nicolas Ballas.
\newblock Revisiting Feature Prediction for Learning Visual Representations from Video, 2024.
\newblock arXiv preprint \url{https://arxiv.org/abs/2404.08471}.

\bibitem[Bousseljot et~al.(1995)Bousseljot, Kreiseler, and Schnabel]{bousseljot1995nutzung}
Ralf Bousseljot, Dieter Kreiseler, and Allard Schnabel.
\newblock Nutzung der EKG-Signaldatenbank CARDIODAT der PTB {\"u}ber das Internet.
\newblock \emph{Biomedizinische Technik / Biomedical Engineering}, 40\penalty0 (s1):\penalty0 317--318, 1995.

\bibitem[Brown et~al.(2020)Brown, Mann, Ryder, Subbiah, Kaplan, Dhariwal, Neelakantan, Shyam, Sastry, Askell, Agarwal, Herbert-Voss, Krueger, Henighan, Child, Ramesh, Ziegler, Wu, Winter, Hesse, Chen, Sigler, Litwin, Gray, Chess, Clark, Berner, McCandlish, Radford, Sutskever, and Amodei]{brown2020language}
Tom Brown, Benjamin Mann, Nick Ryder, Melanie Subbiah, Jared~D Kaplan, Prafulla Dhariwal, Arvind Neelakantan, Pranav Shyam, Girish Sastry, Amanda Askell, Sandhini Agarwal, Ariel Herbert-Voss, Gretchen Krueger, Tom Henighan, Rewon Child, Aditya Ramesh, Daniel Ziegler, Jeffrey Wu, Clemens Winter, Chris Hesse, Mark Chen, Eric Sigler, Mateusz Litwin, Scott Gray, Benjamin Chess, Jack Clark, Christopher Berner, Sam McCandlish, Alec Radford, Ilya Sutskever, and Dario Amodei.
\newblock Language Models are Few-Shot Learners.
\newblock In , \emph{Advances in Neural Information Processing Systems}, volume~33, pages 1877--1901. Curran Associates, Inc., 2020.

\bibitem[Caron et~al.(2018)Caron, Bojanowski, Joulin, and Douze]{caron2018deep}
Mathilde Caron, Piotr Bojanowski, Armand Joulin, and Matthijs Douze.
\newblock Deep Clustering for Unsupervised Learning of Visual Features.
\newblock In \emph{Proceedings of the European Conference on Computer Vision (ECCV)}, 2018.

\bibitem[Caron et~al.(2020)Caron, Misra, Mairal, Goyal, Bojanowski, and Joulin]{caron2020unsupervised}
Mathilde Caron, Ishan Misra, Julien Mairal, Priya Goyal, Piotr Bojanowski, and Armand Joulin.
\newblock Unsupervised Learning of Visual Features by Contrasting Cluster Assignments.
\newblock In , \emph{Advances in Neural Information Processing Systems}, volume~33, pages 9912--9924. Curran Associates, Inc., 2020.

\bibitem[Caron et~al.(2021)Caron, Touvron, Misra, J\'egou, Mairal, Bojanowski, and Joulin]{caron2021emerging}
Mathilde Caron, Hugo Touvron, Ishan Misra, Herv\'e J\'egou, Julien Mairal, Piotr Bojanowski, and Armand Joulin.
\newblock Emerging Properties in Self-Supervised Vision Transformers.
\newblock In \emph{Proceedings of the IEEE/CVF International Conference on Computer Vision (ICCV)}, pages 9650--9660, 2021.

\bibitem[Chen et~al.(2020)Chen, Kornblith, Norouzi, and Hinton]{chen2020simple}
Ting Chen, Simon Kornblith, Mohammad Norouzi, and Geoffrey Hinton.
\newblock A Simple Framework for Contrastive Learning of Visual Representations.
\newblock In , \emph{Proceedings of the 37th International Conference on Machine Learning}, volume 119 of \emph{Proceedings of Machine Learning Research}, pages 1597--1607. PMLR, 2020.

\bibitem[Chen and He(2021)]{chen2021exploring}
Xinlei Chen and Kaiming He.
\newblock Exploring Simple Siamese Representation Learning.
\newblock In \emph{Proceedings of the IEEE/CVF Conference on Computer Vision and Pattern Recognition (CVPR)}, pages 15750--15758, 2021.

\bibitem[{Common Crawl}(2024)]{commoncrawl}
{Common Crawl}.
\newblock {Common Crawl Dataset}.
\newblock \url{https://commoncrawl.org}, 2024.
\newblock [Accessed: August 21, 2024].

\bibitem[Darcet et~al.(2024)Darcet, Oquab, Mairal, and Bojanowski]{darcet2024vision}
Timoth{\'e}e Darcet, Maxime Oquab, Julien Mairal, and Piotr Bojanowski.
\newblock Vision Transformers Need Registers.
\newblock In \emph{The Twelfth International Conference on Learning Representations}, 2024.

\bibitem[Dosovitskiy et~al.(2021)Dosovitskiy, Beyer, Kolesnikov, Weissenborn, Zhai, Unterthiner, Dehghani, Minderer, Heigold, Gelly, Uszkoreit, and Houlsby]{dosovitskiy2021an}
Alexey Dosovitskiy, Lucas Beyer, Alexander Kolesnikov, Dirk Weissenborn, Xiaohua Zhai, Thomas Unterthiner, Mostafa Dehghani, Matthias Minderer, Georg Heigold, Sylvain Gelly, Jakob Uszkoreit, and Neil Houlsby.
\newblock An Image is Worth 16x16 Words: Transformers for Image Recognition at Scale.
\newblock In \emph{International Conference on Learning Representations}, 2021.

\bibitem[Gow et~al.(2023)Gow, Pollard, Nathanson, Johnson, Moody, Fernandes, Greenbaum, Waks, Eslami, Carbonati, Chaudhari, Herbst, Moukheiber, Berkowitz, Mark, and Horng]{gow2023mimic}
Brian Gow, Tom Pollard, Larry~A Nathanson, Alistair Johnson, Benjamin Moody, Chrystinne Fernandes, Nathaniel Greenbaum, Jonathan~W Waks, Parastou Eslami, Tanner Carbonati, Ashish Chaudhari, Elizabeth Herbst, Dana Moukheiber, Seth Berkowitz, Roger Mark, and Steven Horng.
\newblock {MIMIC-IV-ECG}: Diagnostic electrocardiogram matched subset.
\newblock \emph{PhysioNet}, 2023.
\newblock \doi{https://doi.org/10.13026/4nqg-sb35}.

\bibitem[Grill et~al.(2020)Grill, Strub, Altch\'{e}, Tallec, Richemond, Buchatskaya, Doersch, Avila~Pires, Guo, Gheshlaghi~Azar, Piot, kavukcuoglu, Munos, and Valko]{grill2020bootstrap}
Jean-Bastien Grill, Florian Strub, Florent Altch\'{e}, Corentin Tallec, Pierre Richemond, Elena Buchatskaya, Carl Doersch, Bernardo Avila~Pires, Zhaohan Guo, Mohammad Gheshlaghi~Azar, Bilal Piot, koray kavukcuoglu, Remi Munos, and Michal Valko.
\newblock Bootstrap Your Own Latent - A New Approach to Self-Supervised Learning.
\newblock In , \emph{Advances in Neural Information Processing Systems}, volume~33, pages 21271--21284. Curran Associates, Inc., 2020.

\bibitem[He et~al.(2016)He, Zhang, Ren, and Sun]{he2016deep}
Kaiming He, Xiangyu Zhang, Shaoqing Ren, and Jian Sun.
\newblock Deep Residual Learning for Image Recognition.
\newblock In \emph{Proceedings of the IEEE Conference on Computer Vision and Pattern Recognition (CVPR)}, 2016.

\bibitem[He et~al.(2020)He, Fan, Wu, Xie, and Girshick]{he2020momentum}
Kaiming He, Haoqi Fan, Yuxin Wu, Saining Xie, and Ross Girshick.
\newblock Momentum Contrast for Unsupervised Visual Representation Learning.
\newblock In \emph{2020 IEEE/CVF Conference on Computer Vision and Pattern Recognition (CVPR)}, pages 9726--9735, 2020.

\bibitem[He et~al.(2022)He, Chen, Xie, Li, Doll\'ar, and Girshick]{he2022masked}
Kaiming He, Xinlei Chen, Saining Xie, Yanghao Li, Piotr Doll\'ar, and Ross Girshick.
\newblock Masked Autoencoders Are Scalable Vision Learners.
\newblock In \emph{Proceedings of the IEEE/CVF Conference on Computer Vision and Pattern Recognition (CVPR)}, pages 16000--16009, 2022.

\bibitem[He et~al.(2019)He, Zhang, Zhang, Zhang, Xie, and Li]{he2019bag}
Tong He, Zhi Zhang, Hang Zhang, Zhongyue Zhang, Junyuan Xie, and Mu~Li.
\newblock Bag of Tricks for Image Classification with Convolutional Neural Networks.
\newblock In \emph{Proceedings of the IEEE/CVF Conference on Computer Vision and Pattern Recognition (CVPR)}, 2019.

\bibitem[Hochreiter and Schmidhuber(1997)]{hochreiter1997long}
Sepp Hochreiter and J\"{u}rgen Schmidhuber.
\newblock Long Short-Term Memory.
\newblock \emph{Neural Comput.}, 9\penalty0 (8):\penalty0 1735–1780, 1997.
\newblock ISSN 0899-7667.
\newblock \doi{10.1162/neco.1997.9.8.1735}.

\bibitem[Ismail~Fawaz et~al.(2020)Ismail~Fawaz, Lucas, Forestier, Pelletier, Schmidt, Weber, Webb, Idoumghar, Muller, and Petitjean]{ismailfawaz2020inceptiontime}
Hassan Ismail~Fawaz, Benjamin Lucas, Germain Forestier, Charlotte Pelletier, Daniel~F. Schmidt, Jonathan Weber, Geoffrey~I. Webb, Lhassane Idoumghar, Pierre-Alain Muller, and Fran{\c{c}}ois Petitjean.
\newblock InceptionTime: Finding AlexNet for time series classification.
\newblock \emph{Data Mining and Knowledge Discovery}, 34\penalty0 (6):\penalty0 1936--1962, 2020.
\newblock ISSN 1573-756X.
\newblock \doi{10.1007/s10618-020-00710-y}.

\bibitem[Jia et~al.(2021)Jia, Yang, Xia, Chen, Parekh, Pham, Le, Sung, Li, and Duerig]{jia2021scaling}
Chao Jia, Yinfei Yang, Ye~Xia, Yi-Ting Chen, Zarana Parekh, Hieu Pham, Quoc Le, Yun-Hsuan Sung, Zhen Li, and Tom Duerig.
\newblock Scaling Up Visual and Vision-Language Representation Learning With Noisy Text Supervision.
\newblock In , \emph{Proceedings of the 38th International Conference on Machine Learning}, volume 139 of \emph{Proceedings of Machine Learning Research}, pages 4904--4916. PMLR, 2021.

\bibitem[Kachuee et~al.(2018)Kachuee, Fazeli, and Sarrafzadeh]{kachuee2018ecg}
Mohammad Kachuee, Shayan Fazeli, and Majid Sarrafzadeh.
\newblock ECG Heartbeat Classification: A Deep Transferable Representation.
\newblock In \emph{2018 IEEE International Conference on Healthcare Informatics (ICHI)}, pages 443--444, 2018.
\newblock \doi{10.1109/ICHI.2018.00092}.

\bibitem[Kingma and Welling(2013)]{kingma2013auto}
Diederik~P Kingma and Max Welling.
\newblock Auto-encoding variational bayes.
\newblock \emph{arXiv preprint arXiv:1312.6114}, 2013.

\bibitem[Kiyasseh et~al.(2021)Kiyasseh, Zhu, and Clifton]{kiyasseh2021clocs}
Dani Kiyasseh, Tingting Zhu, and David~A Clifton.
\newblock CLOCS: Contrastive Learning of Cardiac Signals Across Space, Time, and Patients.
\newblock In , \emph{Proceedings of the 38th International Conference on Machine Learning}, volume 139 of \emph{Proceedings of Machine Learning Research}, pages 5606--5615. PMLR, 2021.

\bibitem[Kolesnikov et~al.(2020)Kolesnikov, Beyer, Zhai, Puigcerver, Yung, Gelly, and Houlsby]{kolesnikov2020big}
Alexander Kolesnikov, Lucas Beyer, Xiaohua Zhai, Joan Puigcerver, Jessica Yung, Sylvain Gelly, and Neil Houlsby.
\newblock Big Transfer (BiT): General Visual Representation Learning.
\newblock In \emph{Computer Vision – ECCV 2020: 16th European Conference, Glasgow, UK, August 23–28, 2020, Proceedings, Part V}, page 491–507, Berlin, Heidelberg, 2020. Springer-Verlag.
\newblock ISBN 978-3-030-58557-0.
\newblock \doi{10.1007/978-3-030-58558-7_29}.

\bibitem[Lai et~al.(2023)Lai, Tan, Wang, Ji, Guo, Han, Shi, Feng, and Yang]{lai2023practical}
Jiewei Lai, Huixin Tan, Jinliang Wang, Lei Ji, Jun Guo, Baoshi Han, Yajun Shi, Qianjin Feng, and Wei Yang.
\newblock Practical intelligent diagnostic algorithm for wearable 12-lead ECG via self-supervised learning on large-scale dataset.
\newblock \emph{Nature Communications}, 14\penalty0 (1):\penalty0 3741, 2023.
\newblock ISSN 2041-1723.
\newblock \doi{10.1038/s41467-023-39472-8}.

\bibitem[Le et~al.(2023)Le, Truong, Brijesh, Adjeroh, and Le]{le2023sclst}
Duc Le, Sang Truong, Patel Brijesh, Donald~A. Adjeroh, and Ngan Le.
\newblock sCL-ST: Supervised Contrastive Learning With Semantic Transformations for Multiple Lead ECG Arrhythmia Classification.
\newblock \emph{IEEE Journal of Biomedical and Health Informatics}, 27\penalty0 (6):\penalty0 2818--2828, 2023.
\newblock \doi{10.1109/JBHI.2023.3246241}.

\bibitem[Li et~al.(2024)Li, Liu, Cheng, Arcucci, and Hong]{li2024frozen}
Jun Li, Che Liu, Sibo Cheng, Rossella Arcucci, and Shenda Hong.
\newblock Frozen Language Model Helps ECG Zero-Shot Learning.
\newblock In , \emph{Medical Imaging with Deep Learning}, volume 227 of \emph{Proceedings of Machine Learning Research}, pages 402--415. PMLR, 2024.

\bibitem[Liu et~al.(2018)Liu, Liu, Zhao, Zhang, Wu, Xu, Liu, Ma, Wei, He, et~al.]{liu2018open}
Feifei Liu, Chengyu Liu, Lina Zhao, Xiangyu Zhang, Xiaoling Wu, Xiaoyan Xu, Yulin Liu, Caiyun Ma, Shoushui Wei, Zhiqiang He, et~al.
\newblock An open access database for evaluating the algorithms of electrocardiogram rhythm and morphology abnormality detection.
\newblock \emph{Journal of Medical Imaging and Health Informatics}, 8\penalty0 (7):\penalty0 1368--1373, 2018.

\bibitem[Loshchilov and Hutter(2017)]{loshchilov2017sgdr}
Ilya Loshchilov and Frank Hutter.
\newblock {SGDR}: Stochastic Gradient Descent with Warm Restarts.
\newblock In \emph{International Conference on Learning Representations}, 2017.

\bibitem[Loshchilov and Hutter(2019)]{loshchilov2018decoupled}
Ilya Loshchilov and Frank Hutter.
\newblock Decoupled Weight Decay Regularization.
\newblock In \emph{International Conference on Learning Representations}, 2019.

\bibitem[Mehari and Strodthoff(2022)]{mehari2022selfsupervised}
Temesgen Mehari and Nils Strodthoff.
\newblock Self-supervised representation learning from 12-lead ECG data.
\newblock \emph{Computers in Biology and Medicine}, 141:\penalty0 105114, 2022.
\newblock ISSN 0010-4825.
\newblock \doi{https://doi.org/10.1016/j.compbiomed.2021.105114}.

\bibitem[Na et~al.(2024)Na, Park, Tae, and Joo]{na2024guiding}
Yeongyeon Na, Minje Park, Yunwon Tae, and Sunghoon Joo.
\newblock Guiding Masked Representation Learning to Capture Spatio-Temporal Relationship of Electrocardiogram.
\newblock In \emph{The Twelfth International Conference on Learning Representations}, 2024.

\bibitem[Oh et~al.(2022)Oh, Chung, Kwon, Hong, and Choi]{oh2022leadagnostic}
Jungwoo Oh, Hyunseung Chung, Joon-myoung Kwon, Dong-gyun Hong, and Edward Choi.
\newblock Lead-agnostic Self-supervised Learning for Local and Global Representations of Electrocardiogram.
\newblock In , \emph{Proceedings of the Conference on Health, Inference, and Learning}, volume 174 of \emph{Proceedings of Machine Learning Research}, pages 338--353. PMLR, 2022.

\bibitem[Oquab et~al.(2024)Oquab, Darcet, Moutakanni, Vo, Szafraniec, Khalidov, Fernandez, HAZIZA, Massa, El-Nouby, Assran, Ballas, Galuba, Howes, Huang, Li, Misra, Rabbat, Sharma, Synnaeve, Xu, Jegou, Mairal, Labatut, Joulin, and Bojanowski]{oquab2024dinov}
Maxime Oquab, Timoth{\'e}e Darcet, Th{\'e}o Moutakanni, Huy~V. Vo, Marc Szafraniec, Vasil Khalidov, Pierre Fernandez, Daniel HAZIZA, Francisco Massa, Alaaeldin El-Nouby, Mido Assran, Nicolas Ballas, Wojciech Galuba, Russell Howes, Po-Yao Huang, Shang-Wen Li, Ishan Misra, Michael Rabbat, Vasu Sharma, Gabriel Synnaeve, Hu~Xu, Herve Jegou, Julien Mairal, Patrick Labatut, Armand Joulin, and Piotr Bojanowski.
\newblock {DINO}v2: Learning Robust Visual Features without Supervision.
\newblock \emph{Transactions on Machine Learning Research}, 2024.
\newblock ISSN 2835-8856.

\bibitem[Qin et~al.(2023)Qin, Sun, Chen, Yang, Zhang, Fei, and Wang]{qin2023mvkt}
Yuzhen Qin, Li~Sun, Hui Chen, Wenming Yang, Wei-Qiang Zhang, Jintao Fei, and Guijin Wang.
\newblock MVKT-ECG: Efficient single-lead ECG classification for multi-label arrhythmia by multi-view knowledge transferring.
\newblock \emph{Computers in Biology and Medicine}, 166:\penalty0 107503, 2023.
\newblock ISSN 0010-4825.
\newblock \doi{https://doi.org/10.1016/j.compbiomed.2023.107503}.

\bibitem[Radford et~al.(2023)Radford, Kim, Xu, Brockman, McLeavey, and Sutskever]{radford2023robust}
Alec Radford, Jong~Wook Kim, Tao Xu, Greg Brockman, Christine McLeavey, and Ilya Sutskever.
\newblock Robust Speech Recognition via Large-Scale Weak Supervision.
\newblock In \emph{International Conference on Machine Learning}, pages 28492--28518. PMLR, 2023.

\bibitem[Reyna et~al.(2021)Reyna, Sadr, Alday, Gu, Shah, Robichaux, Rad, Elola, Seyedi, Ansari, et~al.]{reyna2021will}
Matthew~A Reyna, Nadi Sadr, Erick A~Perez Alday, Annie Gu, Amit~J Shah, Chad Robichaux, Ali~Bahrami Rad, Andoni Elola, Salman Seyedi, Sardar Ansari, et~al.
\newblock Will {T}wo {D}o? {V}arying {D}imensions in {E}lectrocardiography: the {P}hysioNet/{C}omputing in {C}ardiology {C}hallenge 2021.
\newblock In \emph{2021 Computing in Cardiology (CinC)}, volume~48, pages 1--4. IEEE, 2021.

\bibitem[Ribeiro et~al.(2021)Ribeiro, Paixao, Lima, Horta~Ribeiro, Pinto~Filho, Gomes, Oliveira, Meira~Jr, Schon, and Ribeiro]{ribeiro2021code15}
Antônio~H. Ribeiro, Gabriela~M.M. Paixao, Emilly~M. Lima, Manoel Horta~Ribeiro, Marcelo~M. Pinto~Filho, Paulo~R. Gomes, Derick~M. Oliveira, Wagner Meira~Jr, Thömas~B Schon, and Antonio Luiz~P. Ribeiro.
\newblock {CODE-15\%: a large scale annotated dataset of 12-lead ECGs}.
\newblock \emph{Zenodo}, June 2021.
\newblock \doi{10.5281/zenodo.4916206}.

\bibitem[Rieke et~al.(2020)Rieke, Hancox, Li, Milletar{\`i}, Roth, Albarqouni, Bakas, Galtier, Landman, Maier-Hein, Ourselin, Sheller, Summers, Trask, Xu, Baust, and Cardoso]{rieke2020future}
Nicola Rieke, Jonny Hancox, Wenqi Li, Fausto Milletar{\`i}, Holger~R. Roth, Shadi Albarqouni, Spyridon Bakas, Mathieu~N. Galtier, Bennett~A. Landman, Klaus Maier-Hein, S{\'e}bastien Ourselin, Micah Sheller, Ronald~M. Summers, Andrew Trask, Daguang Xu, Maximilian Baust, and M.~Jorge Cardoso.
\newblock The future of digital health with federated learning.
\newblock \emph{npj Digital Medicine}, 3\penalty0 (1):\penalty0 119, 2020.
\newblock ISSN 2398-6352.

\bibitem[Salem et~al.(2018)Salem, Taheri, and Yuan]{salem2018ecg}
Milad Salem, Shayan Taheri, and Jiann–Shiun Yuan.
\newblock ECG Arrhythmia Classification Using Transfer Learning from 2- Dimensional Deep CNN Features.
\newblock In \emph{2018 IEEE Biomedical Circuits and Systems Conference (BioCAS)}, pages 1--4, 2018.
\newblock \doi{10.1109/BIOCAS.2018.8584808}.

\bibitem[Sawano et~al.(2022)Sawano, Kodera, Takeuchi, Sukeda, Katsushika, and Komuro]{sawano2022masked}
Shinnosuke Sawano, Satoshi Kodera, Hirotoshi Takeuchi, Issei Sukeda, Susumu Katsushika, and Issei Komuro.
\newblock Masked Autoencoder-Based Self-Supervised Learning for Electrocardiograms to Detect Left Ventricular Systolic Dysfunction.
\newblock In \emph{NeurIPS 2022 Workshop on Learning from Time Series for Health}, 2022.

\bibitem[Schuhmann et~al.(2021)Schuhmann, Vencu, Beaumont, Kaczmarczyk, Mullis, Katta, Coombes, Jitsev, and Komatsuzaki]{schuhmann2021laion400}
Christoph Schuhmann, Richard Vencu, Romain Beaumont, Robert Kaczmarczyk, Clayton Mullis, Aarush Katta, Theo Coombes, Jenia Jitsev, and Aran Komatsuzaki.
\newblock LAION-400M: Open Dataset of CLIP-Filtered 400 Million Image-Text Pairs, 2021.
\newblock arXiv preprint \url{https://arxiv.org/abs/2111.02114}.

\bibitem[Soltanieh et~al.(2022)Soltanieh, Etemad, and Hashemi]{soltanieh2022analysis}
Sahar Soltanieh, Ali Etemad, and Javad Hashemi.
\newblock Analysis of Augmentations for Contrastive ECG Representation Learning.
\newblock In \emph{2022 International Joint Conference on Neural Networks (IJCNN)}, pages 1--10, 2022.
\newblock \doi{10.1109/IJCNN55064.2022.9892600}.

\bibitem[Strodthoff et~al.(2021)Strodthoff, Wagner, Schaeffter, and Samek]{strodthoff2021deep}
Nils Strodthoff, Patrick Wagner, Tobias Schaeffter, and Wojciech Samek.
\newblock Deep Learning for ECG Analysis: Benchmarks and Insights from PTB-XL.
\newblock \emph{IEEE Journal of Biomedical and Health Informatics}, 25\penalty0 (5):\penalty0 1519--1528, 2021.
\newblock \doi{10.1109/JBHI.2020.3022989}.

\bibitem[Touvron et~al.(2023)Touvron, Martin, Stone, Albert, Almahairi, Babaei, Bashlykov, Batra, Bhargava, Bhosale, et~al.]{touvron2023llama2}
Hugo Touvron, Louis Martin, Kevin Stone, Peter Albert, Amjad Almahairi, Yasmine Babaei, Nikolay Bashlykov, Soumya Batra, Prajjwal Bhargava, Shruti Bhosale, et~al.
\newblock Llama 2: Open foundation and fine-tuned chat models.
\newblock \emph{arXiv preprint arXiv:2307.09288}, 2023.

\bibitem[Vaid et~al.(2023)Vaid, Jiang, Sawant, Lerakis, Argulian, Ahuja, Lampert, Charney, Greenspan, Narula, Glicksberg, and Nadkarni]{vaid2023foundational}
Akhil Vaid, Joy Jiang, Ashwin Sawant, Stamatios Lerakis, Edgar Argulian, Yuri Ahuja, Joshua Lampert, Alexander Charney, Hayit Greenspan, Jagat Narula, Benjamin Glicksberg, and Girish~N. Nadkarni.
\newblock A foundational vision transformer improves diagnostic performance for electrocardiograms.
\newblock \emph{npj Digital Medicine}, 6\penalty0 (1):\penalty0 108, 2023.
\newblock ISSN 2398-6352.
\newblock \doi{10.1038/s41746-023-00840-9}.

\bibitem[van~den Oord et~al.(2019)van~den Oord, Li, and Vinyals]{oord2019representation}
Aaron van~den Oord, Yazhe Li, and Oriol Vinyals.
\newblock Representation Learning with Contrastive Predictive Coding.
\newblock \emph{arXiv preprint arXiv:1807.03748}, 2019.

\bibitem[Vaswani et~al.(2017)Vaswani, Shazeer, Parmar, Uszkoreit, Jones, Gomez, Kaiser, and Polosukhin]{vaswani2017attention}
Ashish Vaswani, Noam Shazeer, Niki Parmar, Jakob Uszkoreit, Llion Jones, Aidan~N Gomez, \L~ukasz Kaiser, and Illia Polosukhin.
\newblock Attention is All you Need.
\newblock In , \emph{Advances in Neural Information Processing Systems}, volume~30. Curran Associates, Inc., 2017.

\bibitem[Vincent et~al.(2008)Vincent, Larochelle, Bengio, and Manzagol]{vincent2008extracting}
Pascal Vincent, Hugo Larochelle, Yoshua Bengio, and Pierre-Antoine Manzagol.
\newblock Extracting and composing robust features with denoising autoencoders.
\newblock In \emph{Proceedings of the 25th International Conference on Machine Learning}, ICML '08, page 1096–1103, New York, NY, USA, 2008. Association for Computing Machinery.
\newblock ISBN 9781605582054.
\newblock \doi{10.1145/1390156.1390294}.

\bibitem[Wagner et~al.(2020)Wagner, Strodthoff, Bousseljot, Kreiseler, Lunze, Samek, and Schaeffter]{wagner2020ptbxl}
Patrick Wagner, Nils Strodthoff, Ralf-Dieter Bousseljot, Dieter Kreiseler, Fatima~I Lunze, Wojciech Samek, and Tobias Schaeffter.
\newblock {PTB-XL}, a large publicly available electrocardiography dataset.
\newblock \emph{Scientific Data}, 7\penalty0 (1):\penalty0 154, 2020.

\bibitem[Wang et~al.(2017)Wang, Yan, and Oates]{wang2017time}
Zhiguang Wang, Weizhong Yan, and Tim Oates.
\newblock Time series classification from scratch with deep neural networks: A strong baseline.
\newblock In \emph{2017 International joint conference on neural networks (IJCNN)}, pages 1578--1585. IEEE, 2017.

\bibitem[Weimann and Conrad(2021)]{weimann2021transfer}
Kuba Weimann and Tim O.~F. Conrad.
\newblock Transfer learning for ECG classification.
\newblock \emph{Scientific Reports}, 11\penalty0 (1):\penalty0 5251, 2021.
\newblock ISSN 2045-2322.
\newblock \doi{10.1038/s41598-021-84374-8}.

\bibitem[Xie et~al.(2022)Xie, Zhang, Cao, Lin, Bao, Yao, Dai, and Hu]{xie2022simmim}
Zhenda Xie, Zheng Zhang, Yue Cao, Yutong Lin, Jianmin Bao, Zhuliang Yao, Qi~Dai, and Han Hu.
\newblock SimMIM: A Simple Framework for Masked Image Modeling.
\newblock In \emph{Proceedings of the IEEE/CVF Conference on Computer Vision and Pattern Recognition (CVPR)}, pages 9653--9663, 2022.

\bibitem[Zhang et~al.(2023)Zhang, Liu, Shi, Chang, Wang, He, and Huang]{zhang2023maefe}
Huaicheng Zhang, Wenhan Liu, Jiguang Shi, Sheng Chang, Hao Wang, Jin He, and Qijun Huang.
\newblock MaeFE: Masked Autoencoders Family of Electrocardiogram for Self-Supervised Pretraining and Transfer Learning.
\newblock \emph{IEEE Transactions on Instrumentation and Measurement}, 72:\penalty0 1--15, 2023.
\newblock \doi{10.1109/TIM.2022.3228267}.

\bibitem[Zheng et~al.(2020{\natexlab{a}})Zheng, Chu, Struppa, Zhang, Yacoub, El-Askary, Chang, Ehwerhemuepha, Abudayyeh, Barrett, Fu, Yao, Li, Guo, and Rakovski]{zheng2020optimal}
Jianwei Zheng, Huimin Chu, Daniele Struppa, Jianming Zhang, Sir~Magdi Yacoub, Hesham El-Askary, Anthony Chang, Louis Ehwerhemuepha, Islam Abudayyeh, Alexander Barrett, Guohua Fu, Hai Yao, Dongbo Li, Hangyuan Guo, and Cyril Rakovski.
\newblock Optimal Multi-Stage Arrhythmia Classification Approach.
\newblock \emph{Scientific Reports}, 10\penalty0 (1):\penalty0 2898, 2020{\natexlab{a}}.
\newblock ISSN 2045-2322.
\newblock \doi{10.1038/s41598-020-59821-7}.

\bibitem[Zheng et~al.(2020{\natexlab{b}})Zheng, Zhang, Danioko, Yao, Guo, and Rakovski]{Zheng2020a12lead}
Jianwei Zheng, Jianming Zhang, Sidy Danioko, Hai Yao, Hangyuan Guo, and Cyril Rakovski.
\newblock A 12-lead electrocardiogram database for arrhythmia research covering more than 10,000 patients.
\newblock \emph{Scientific Data}, 7\penalty0 (1):\penalty0 48, 2020{\natexlab{b}}.
\newblock ISSN 2052-4463.
\newblock \doi{10.1038/s41597-020-0386-x}.

\end{thebibliography}
